\documentclass[fleqn,usenatbib,useAMS]{mnras}

\usepackage{graphicx}	
\usepackage{amsmath}	
\usepackage{amssymb}	
\usepackage{multicol}        
\usepackage{bm}		
\usepackage{pdflscape}	

\usepackage[T1]{fontenc}
\usepackage{ae,aecompl}

\usepackage{newtxtext,newtxmath}

\title[Modeling SN 2005cl with RSG progenitors]{Modeling the light curve of Type IIn-P SN 2005cl with red supergiant progenitors featuring pre-SN ourbursts}

\author[C. Li \& V. Morozova]{Chunhui Li,$^{1}$%
\thanks{Contact e-mail: \href{mailto:cpl5430@psu.edu}{cpl5430@psu.edu}}%
Viktoriya Morozova$^{1,2}$%
\thanks{Contact e-mail: \href{mailto:vzg51380@psu.edu}{vzg5138@psu.edu}}%
\\
$^{1}$Department of Physics, The Pennsylvania State University, University Park, PA 16802-6300, USA \\
$^{2}$Institute for Gravitation and the Cosmos, The Pennsylvania State University, University Park, PA 16802-6300, USA}

\pubyear{2021}

\begin{document}
\label{firstpage}
\pagerange{\pageref{firstpage}--\pageref{lastpage}}
\maketitle

\begin{abstract}
All Type IIn supernovae (SNe IIn) show narrow hydrogen emission lines in their spectra. Apart from this common feature, they demonstrate very broad diversity in brightness, duration, and morphology of their light curves, which indicates that they likely come from a variety of progenitor systems and explosion channels. A particular subset of SNe IIn, the so called SNe IIn-P, exhibit $\sim$100 days plateau phases that are very similar to the ones of the ordinary hydrogen-rich SNe (SNe II). In the past, SNe IIn-P were explained by the models of sub-energetic electron capture explosions surrounded by dense extended winds. In this work, we attempt to explain this class of SNe with standard red supergiant (RSG) progenitors that experience outbursts several month before the final explosion. The outburst energies that show the best agreement between our models and the data ($5\times10^{46}\,{\rm erg}$) fall at the low range of the outburst energies that have been observed for SNe IIn (between few times $10^{46}\,{\rm erg}$ and $10^{49}\,{\rm erg}$). Instead, the inferred explosion energy of SN 2005cl is relatively high ($1-2\times10^{51}\,{\rm erg}$) compared to the explosion energies of the ordinary SNe II. Our models provide alternative explanation of SNe IIn-P to the previously proposed scenarios.
\end{abstract}





\section{Introduction}

Hydrogen-rich supernovae (SNe) that exhibit narrow Balmer emission lines in their spectra are commonly distinguished as a separate subclass of SNe, the so called Type IIn \citep{schlegel:1990}. SNe IIn are often brighter than regular hydrogen-rich SNe and exhibit more extended light curves \citep{kiewe:2012,nyholm:2020}. It has long been understood that the narrow lines in their spectra originate from the interaction of shocked ejecta with the slow and dense material located ahead of the shock \citep{chugai:1991,chugai:1994,chugai:2001,chugai:2004}. Moreover, it has been recognized that almost any astrophysical explosion, such as core-collapse or thermonuclear SN, may appear as a SN IIn if surrounded by a sufficient amount of the circumstellar material (CSM) \citep{smith:2017}. Indeed, the broad diversity of light curves that can be classified as Type IIn suggests that they may originate from very different progenitors \citep[see, for example,][]{taddia:2013,fox:2013,habergham:2014,delarosa:2016}.

A variety of mechanisms could lead to the formation of this CSM, ranging from enhanced pre-SN winds to the binary interactions \citep{moriya:2014,moriya:2014a,andrews:2017,kurfurst:2020}. One of the most promising scenarios is that this material forms as a result of a violent mass eruption due to the core nuclear burning in the last years before the SN explosion \citep{quataert:2012,shiode:2014,dessart:2016,Fuller:2017,das:2017,leung:2020,takei:2021,wu:2021}. This idea is supported by numerous observations of SN precursors, outbursts that happen months to years prior to SNe IIn \citep{ofek:2013,mauerhan:2013,fraser:2013,margutti:2014,tartaglia:2016,elias-rosa:2016,ofek:2016,thoene:2017,nyholm:2017,pastorello:2018,reguitti:2019}. Some of the largest collections of SN IIn precursors can be found in \citet{ofek:2014} and \citet{strotjohann:2021}.

Contrary to SNe IIn, the ordinary hydrogen-rich plateau SNe (SNe II) were thought to originate from red supergiant (RSG) progenitors surrounded by regular low-density stellar winds. Only in recent years the early observations of SNe II started to indicate that their progenitors may also be surrounded by the dense CSM \citep{galyam:2014,smith:2015,khazov:2016,yaron:2017,hosseinzadeh:2018,nakaoka:2018,bullivant:2018}. This is supported by hydrodynamical simulations of their light curves, which show significantly better agreement with the observational data when the dense CSM is added to the models \citep{nagy:2016,morozova:2017,morozova:2017a,paxton:2018,moriya:2018,foerster:2018}. In addition, the analysis of late-time nuclear burning in RSG cores shows that they are capable of generating the outbursts with sufficient energy \citep{Fuller:2017}. The fact that no precursors prior to regular SNe II have been detected yet \citep{kochanek:2017,johnson:2017,oneill:2018} may indicate that they are weaker and shorter living than the SN IIn precursors.

That said, there potentially exists a group of SNe that originates from common RSG progenitors, spans across the II and IIn classes, and can be described by a continuously varying characteristics of a pre-SN outburst (energy or timing). Indeed, the observed progenitors of some SNe IIn have moderate masses of $10\,M_{\odot}$ which are typical of RSGs \citep{prieto:2008,szczygiel:2012}. Moreover, \citet{mauerhan:2013a} distinguishes a subclass of SNe (SNe IIn-P) that exhibit narrow lines in their spectra, yet demonstrate features that are typical of regular plateau SNe \citep[see also][]{smith:2017}.

In this work, we use radiation hydrodynamics simulations to probe the idea that SNe IIn-P may come from regular RSG progenitors that experience eruptive outbursts before their explosions. We construct a grid of theoretical light curves that differ by the outburst energy, the final SN energy, and the time interval between the outburst and the SN explosion. With this grid, we look for the best fitting model of SN 2005cl \citep{kiewe:2012}, which is a member of IIn-P class. We confirm that this SN could originate from a RSG progenitor and give a rough estimate of the outburst and explosion parameters for the class.

The rest of this paper is organized as follows. In Section~\ref{setup}, we describe our numerical setup. Section~\ref{results} contains our main results, and Section~\ref{summary} is devoted to the conclusions and discussion.

\section{Numerical Setup}
\label{setup}

We carry out our research using the publicly available code SNEC \citep{Morozova:2015}. SNEC solves the Lagrangian hydrodynamics equations coupled with radiation transport in the flux-limited diffusion approximation. This code is well suited for modeling bolometric light curves of SNe II starting from few days after the shock breakout until the end of the plateau phase. We work with two RSG stellar evolution models at the onset of core-collapse, both taken from the KEPLER set by \citet{sukhbold:2016}. At zero-age main sequence (ZAMS) these models had masses $10\,M_{\odot}$ and $15\,M_{\odot}$. The final mass of the $10\,M_{\odot}$ RSG before the core collapse is $9.7\,M_{\odot}$, and its radius is $513\,R_{\odot}$. The final mass of the $15\,M_{\odot}$ RSG before the core collapse is $12.6\, M_{\odot}$, and its radius is $841\, R_{\odot}$.  

In the first stage of our modeling, we inject a certain amount of energy, $E_{\rm inj}$, at the base of the hydrogen envelope of the RSG. This step is meant to simulate a weak outburst caused by
a convectively driven wave during vigorous late-stage nuclear burning in the RSG core \citep{Fuller:2017}. To inject the energy into the $10\,M_{\odot}$ model, we first excise its core down to the density $0.89\, {\rm g}\,{\rm cm}^{-3}$ ($2.13\,M_{\odot}$ in terms of excised inner mass). In the $15\,M_{\odot}$ model, we excise the core down to the density $1.13\,{\rm g}\,{\rm cm}^{-3}$ ($4.31\, M_{\odot}$ in terms of excised inner mass)\footnote{The amount of mass that we excise at this stage does not carry any physical meaning, but it is determined solely by the balance of factors influencing the feasibility of our numerical simulations. On the one hand, we want the inner boundary of our model to be deeper than the interface between He core and H envelope where we inject the energy. On the other hand, not excising the core leads to slow simulations and numerical problems at the center, since SNEC is not a stellar evolution code and it is not meant to support RSG cores for an extended period of time.}. We utilize the thermal bomb mechanism and inject the energy in the zone with density $0.52\times10^{-4}\, {\rm g}\,{\rm cm}^{-3}$, at the interface between He core and H envelope. The injected energy, $E_{\rm inj}$, takes the values $0.5$, $1.0$, $2.0$, $3.0$ and $4.0\times10^{47}\,{\rm erg}$.


The weak shock wave caused by the energy injection takes about $100$ days to propagate out to the surface of the RSG and preheat its envelope. Once it reaches the surface, it causes ejection of the outermost material and expansion of the star. The amount of the ejected material depends on the energy $E_{\rm inj}$. We follow the evolution of the model for up to $\sim2$ years after the energy injection and collect the snapshots of its profile at different times, $t_{\rm inj}$, for the subsequent explosion with typical SN energies. That is, $t_{\rm inj}$ in our models is the time between the energy injection at the base of the H envelope and the SN explosion.


In the second stage of our modeling, we process the collected profiles in SNEC using regular core-collapse SN setup \citep{Morozova:2015}. Before exploding the models as regular SNe, we attach back their core parts that were excised earlier, assuming that they have not been affected by the weak energy injection. Then, we deposit the explosion energy into the innermost zones of the models, now excising only $1.4\,M_{\odot}$ of material, which is meant to form a stable neutron star remnant. We use four values for the final energy\footnote{In SNEC, $E_{\rm fin}=E_{\rm init}+E_{\rm bomb}$ represents the final energy of the model, where $E_{\rm init}$ is its total initial energy, and $E_{\rm bomb}$ is the energy deposited in the thermal bomb.}, $E_{\rm fin}=0.5$, $1.0$, $1.5$, and $2.0\times10^{51}\,{\rm erg}$. As a result, we obtain a coarse three-dimensional grid of SN light curves differing by $E_{\rm inj}$, $t_{\rm inj}$, and $E_{\rm fin}$. Parsing though this grid we look for the light curve that fits best the data of SN 2005cl by minimizing $\chi^2$ between the data and the theoretical light curves. A very similar approach has been adopted in our earlier works, \citet{morozova:2020} and \citet{tinyanont:2021} \citep[see also][]{grasberg:1986,grasberg:1991}.

One more parameter that has to be chosen in SNEC simulations is the mass of radioactive $^{56}$Ni responsible for the radioactive tail of the light curve. In SNe IIn the $^{56}$Ni mass is not constrained as well as in regular SNe II, and we did not find the estimate for the $^{56}$Ni mass of SN 2005cl in the literature. For this reason, in our simulation we used an average $^{56}$Ni mass typical for SNe II, $M_{\rm Ni}=0.05\,M_{\odot}$ (in SNe II, the $^{56}$Ni masses range between $\sim0.001\,M_{\odot}$ and $\sim0.1\,M_{\odot}$, somewhat clustering around $0.05\,M_{\odot}$; see \citealt{valenti:2016}). When comparing the resulting light curves to the data (in the next section), we see that this value slightly overestimates the observed brightness of the tail, but not by much. In addition, SNEC is less accurate in modeling the radioactive tails of SN light curves compared to their plateau parts, because the black body approximation is no longer suitable after the end of the plateau. For now, we adopt this value of $M_{\rm Ni}$ assuming that it does not strongly affect the deduction of parameters that are of our main interest here ($E_{\rm inj}$, $t_{\rm inj}$, and $E_{\rm fin}$).

\section{Results}
\label{results}

\begin{figure}
    \includegraphics[width=0.48\textwidth]{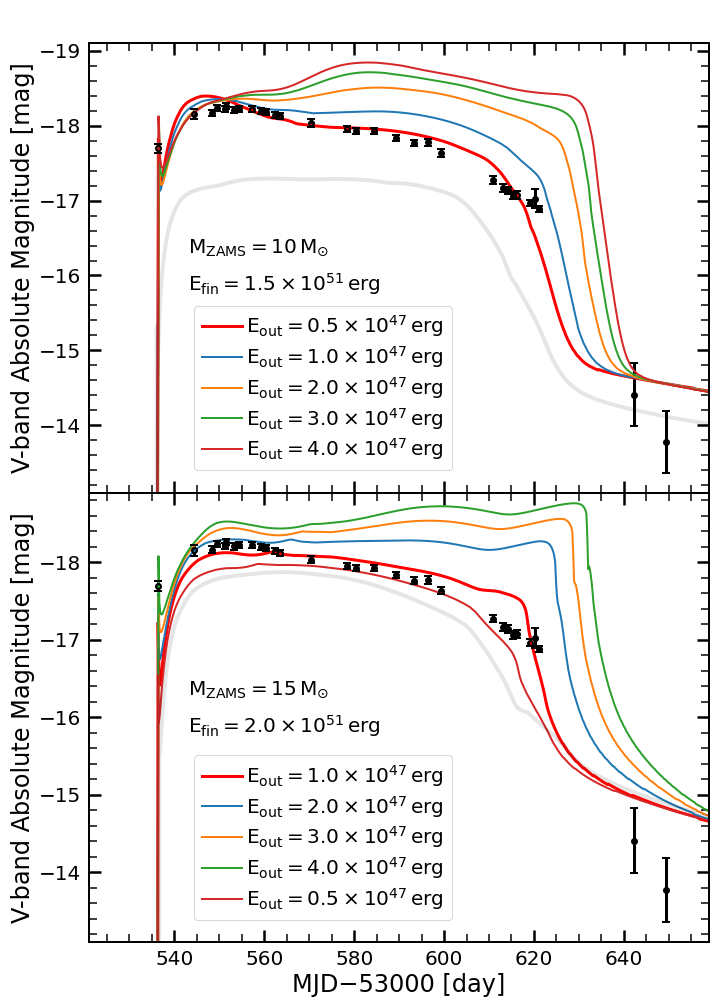}

    \caption{The V-band light curves of the models preheated with different energies $E_{\rm inj}$, compared to the data of SN 2005cl. The top panel shows the $10\,M_{\odot}$ progenitor, and the bottom panel shows the $15\,M_{\odot}$ progenitor. All curves shown in the top panel correspond to same final energy $E_{\rm fin}=1.5\times 10^{51}\,{\rm erg}$ and time $t_{\rm inj}=258$ days. All curves from the bottom panel have the final energy $E_{\rm fin}=2.0\times 10^{51}\,{\rm erg}$ and time $t_{\rm inj}=258$ days. The light curves of bare RSG models with the corresponding explosion energies are shown in gray. The red curve in each panel is the best fitting light curve for that progenitor model.}
    \label{figure 1}
\end{figure}

\begin{figure}
    \includegraphics[width=0.48\textwidth]{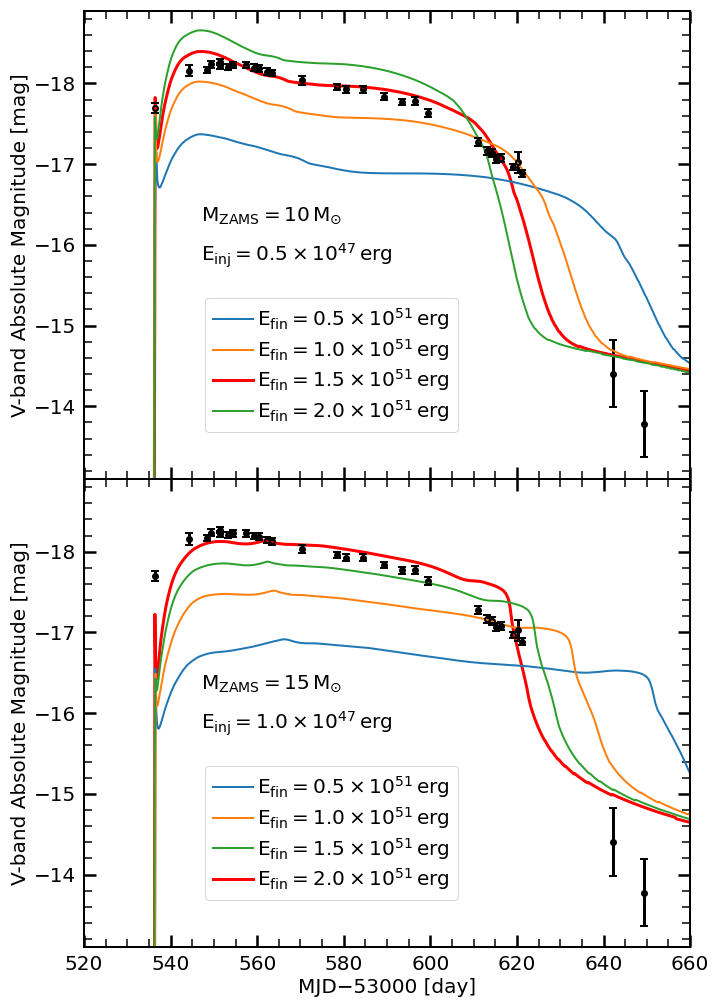}

    \caption{The V-band light curves of the models exploded with different final energies $E_{\rm fin}$, compared to the data of SN 2005cl. The top panel shows the $10\,M_{\odot}$ progenitor, and the bottom panel shows the $15\,M_{\odot}$ progenitor. All curves shown in the top panel correspond to the same outburst energy $E_{\rm out}=0.5\times 10^{47}\,{\rm erg}$ and time $t_{\rm inj}=258$ days. All curves from the bottom panel have the same $E_{\rm out}=1.0\times 10^{47}\,{\rm erg}$ and time $t_{\rm inj}=258$ days. The red curve in each panel is the best fitting light curve for that progenitor model.}
    \label{figure 2}
 \end{figure}

\begin{figure}
    \includegraphics[width=0.48\textwidth]{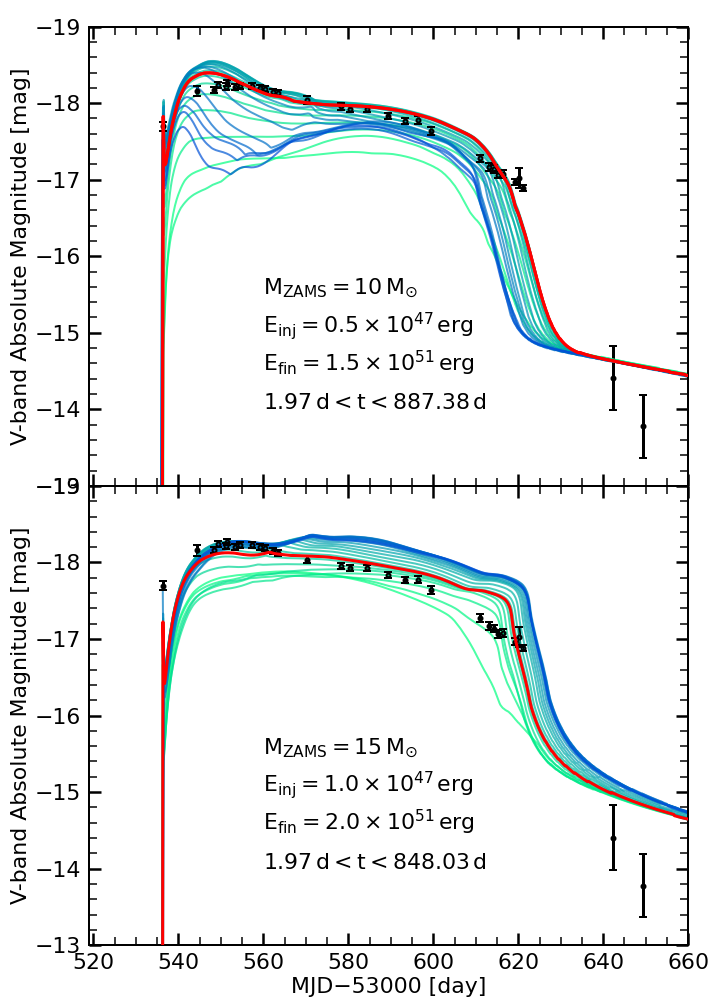}

    \caption{The V-band light curves of the $10\,M_{\odot}$  and $15\,M_{\odot}$ models with different time intervals $t_{\rm inj}$ between the weak outburst and the SN explosion.}
    \label{figure 3}
\end{figure}

Figs.~\ref{figure 1}-\ref{figure 3} show the light curves obtained from our models versus the observed light curve of SN 2005cl in V-band. In all three figures, the red bold lines represent the best fitting light curves for the corresponding RSG progenitor ($10\,M_{\odot}$ or $15\,M_{\odot}$), while the other curves show how the light curve changes when we vary one of the model parameters ($E_{\rm inj}$, $E_{\rm fin}$, or $t_{\rm inj}$). 


In Fig.~\ref{figure 1}, we plot the light curves of $10\,M_{\odot}$ (top panel) and $15\,M_{\odot}$ (bottom panel) RSG models with the final energies $E_{\rm fin}=1.5\times10^{51}\,{\rm erg}$ and $2.0\times10^{51}\,{\rm erg}$, respectively. Different light curves correspond to the different outburst energies $E_{\rm inj}$. The plot demonstrates that the outburst energy has even greater influence on the level and duration of the plateau than it has on the early light curve. Stronger outbursts correspond to markedly brighter and longer plateau phases in both progenitor models. The data show the agreement with relatively modest outburst energies of $5.0\times10^{46}\,{\rm erg}$ for the $10\,M_{\odot}$ progenitor and $10.0\times10^{46}\,{\rm erg}$ for the $15\,M_{\odot}$ progenitor. These values are on the lower end of the range of outburst energies measured by \citet{strotjohann:2021} (between $3$ and $1000\times10^{46}\,{\rm erg}$). 

Almost all light curves in Fig.~\ref{figure 1} show very sharp transition between the plateau phase and the $^{56}$Ni tail which is typical of SNe IIn-P \citep{smith:2017}. Note that we used only one value for the $^{56}$Ni mass, which is close to the average value for SNe II, whereas the observations of SNe IIn-P suggest low $^{56}$Ni yields. Lower values of $^{56}$Ni mass would make the transition even sharper and lead to a better agreement between the data and our models during the tail phase. On the other hand, it would make the end of the plateau dimmer and shorter, which could favor the models with slightly larger outburst energies.

The gray curves in both panels show the light curves obtained from the bare RSG models with the same explosion energies. The plateau brightness of these light curves is insufficient to fit the data, however, the larger explosion energies would make the plateau even shorter. In addition, not preheated RSG models would have more difficulties reproducing the early ($\sim20$ days post explosion) peak and subsequent decline of the observed light curve. This indicates that simulating pre-SN outbursts plays a crucial role in fitting SNe IIn-P light curves by RSG progenitors.


In Fig.~\ref{figure 2}, we plot the light curves of both models for different values of final energies $E_{\rm fin}$, while keeping $E_{\rm inj}$ constant. The plots demonstrate a well-known trend, where an increase in the explosion energy leads to a brighter but shorter plateau \citep{litvinova:1983,kasen:2009,bersten:2011}. The best fitting models have $E_{\rm fin}=1.5\times10^{51}\,{\rm erg}$ for the $10\,M_{\odot}$ progenitor and $E_{\rm fin}=2.0\times10^{51}\,{\rm erg}$ for the $15\,M_{\odot}$ progenitor.

Fig.~\ref{figure 3} shows the light curves obtained from the models with same $E_{\rm inj}$ and $E_{\rm fin}$, but different time intervals between the weak outburst and the SN explosion, $t_{\rm inj}$. Lighter colors correspond to shorter $t_{\rm inj}$, and vice versa. For the small values of $t_{\rm inj}$, the SN explosion happens while the outer layers of the star are still expanding after being preheated, which leads to the brighter light curves, especially during the first $30-40$ days past explosion. However, if $t_{\rm inj}$ is sufficiently large, some of the material ejected in the weak outburst starts to fall back down onto the star. This is clearly seen in Fig.~\ref{figure 3} for the $10\,M_{\odot}$ progenitor (top panel), where the shape of the early light curves becomes different for the models with large $t_{\rm inj}$. The same is not seen for the $15\,M_{\odot}$ progenitor because of the larger outburst energy $E_{\rm inj}$.

The best fitting models for both $10\,M_{\odot}$ and $15\,M_{\odot}$ progenitors have $t_{\rm inj}=258$ days. This agrees with the results of \citet{strotjohann:2021}, where the outbursts were seen anywhere between few tens of days to $\sim700$ days prior to the SN explosions.

\section{Summary and Discussion}
\label{summary}

We have simulated a grid of theoretical light curves from two RSG models preheated by relatively small amount of energy injected at the base of their hydrogen envelopes few days to a few years prior to their explosions as core-collapse SNe. These models are meant to represent SNe preceded by weak outbursts. Varying the outburst energies, the explosion energies, and the time interval between the outburst and the explosion, we found a set of parameters that matches the light curve of SN 2005cl. In our picture, this SN could originate from a relatively low mass, $10\,M_{\odot}$ RSG progenitor that experienced a $5\times10^{46}\,{\rm erg}$ outburst 258 days before it exploded with final energy $1.5\times10^{51}\,{\rm erg}$.

SN 2005cl belongs to the class of SNe IIn-P \citep{mauerhan:2013a,smith:2017}, whose members have persistent narrow hydrogen lines in their spectra yet show the characteristics of ordinary SNe II such as a $\sim$100 days plateau steeply transitioning into a radioactive tail. Among other SNe that belong to the same class are 1994W, 2011ht, and 2009kn. These SNe have similar plateau lengths between $100$ and $125$ days, and they vary in plateau brighness by $\sim$2 magnitudes, with SN 2005cl being the brightest of the four \citep[see][]{mauerhan:2013a}. We speculate that the fits to the other SNe from this group may be found among the models with similar parameters to the one of SN 2005cl. Accurate modeling would require larger number of theoretical light curves than presented here and more precise estimates of the radioactive $^{56}$Ni mass. However, already with the models shown here we can see that the outburst plays a very important role in extending the plateau to the observed length. A bare RSG model exploded with energy needed to reproduce the bright plateau would produce a much shorter light curve lasting only $60-80$ days (see Fig.~\ref{figure 2}).

If the circumstellar material surrounding the SNe before their explosion forms in an eruptive outburst, it would be interesting to compare the amounts of the outburst energy needed to explain the observations of SNe IIn versus regular SNe II. There are relatively few models of SNe II with pre-SN outbursts, but some of the published ones indicate that the outburst energy of $\sim5\times10^{46}\,{\rm erg}$ is sufficient to bring the models in good agreement with observational data \citep{morozova:2020,tinyanont:2021}. The models presented here indicate that the amount of energy needed to explain SNe IIn-P is roughly the same. It is possible to explain the injection of this energy into the base of the hydrogen envelope by the late-time nuclear burning processes happening in the RSG cores \citep{Fuller:2017}.

Similarity of the energies needed for the outbursts in SNe IIn-P and regular SNe II returns us back to the question of why there is still no ordinary SN II with a documented pre-SN outburst. In our opinion, the reason lies in the weakness of the outburst light curve for that amount of energy. Looking at the pre-SN data of SNe with similar outburst energies from \citet{strotjohann:2021}, we see that those outbursts are very hard to recognize with a naked eye, because they are short lived and embedded in noisy data (see, for example, SN 2019bxq with the outburst energy $3\times10^{46}\,{\rm erg}$ released $\sim$350 and $\sim$180 days before the SN, or SN 2018eru with the outburst energy $4\times10^{46}\,{\rm erg}$ released $\sim$200 and $\sim$65 days before the SN). Extracting those outbursts from the data required delicate techniques such as forced photometry and light curve binning \citep{strotjohann:2021}.

On the other hand, if SNe IIn-P and regular SNe II experience similar pre-SN outbursts, why do we not see narrow lines in the spectra of the latter? A possible answer to this question lies in the analysis performed by \citet{strotjohann:2021} of the impact of pre-SN outbursts on the SN spectra. Specifically, the authors showed that almost none of the outbursts detected in the last years before the explosion could be responsible for the formation of narrow lines in the spectra of the subsequent SNe. This is because the radius of material accountable for the narrow lines (as estimated from the spectra) is significantly larger than the radius to which the outburst material could have possibly expanded by that time (as estimated from the timing of the outbursts using a broad range of ejecta velocities). Therefore, it was concluded that earlier episodes of mass eruption that happened years prior to the observed outbursts would be needed to explain the narrow lines seen in the spectra \citep[see also][]{moriya:2014a}. Almost all transients described in \citet{strotjohann:2021} were pre-selected to be of type IIn (except one superluminous SN II, and one SN Ia), because this type was previously associated with the pre-SN activity. However, the outbursts that happen within a year of the explosion do not necessarily cause the long lasting narrow lines, and may instead cause only the so called flash ionization seen in the very early (hours to days) spectra of many SNe II \citep{khazov:2016,yaron:2017,lin:2021}. It would be very interesting to look at a sample of regular well-observed SNe II with sufficient pre-SN data, and check whether or not there are low energy outbursts in the data.

In view of our models, the main parameter that distinguishes SNe IIn-P from ordinary SNe II is the final energy of the explosion. To achieve high plateau brightness seen in SNe IIn one needs the energies of the order of a few times $10^{51}\,{\rm erg}$. On the other hand, the final energies of regular SNe II lie in the range between $0.2$ and $1\times10^{51}\,{\rm erg}$ clustering around $0.5\times10^{51}\,{\rm erg}$ \citep[see, for example,][]{pumo:2017,morozova:2017a}. Interestingly, some of the previous models suggested that SNe IIn-P originate from low energy ($\sim10^{50}\,{\rm erg}$) electron capture explosions \citep[see, for example][]{moriya:2014b,smith:2017}, which would be consistent with low $^{56}$Ni yield seen in these SNe. In our models, lower explosion energies would lead to the dimmer and longer plateau phases incompatible with observations, whereas larger outbursts energies would also increase the length of the plateau. More accurate investigation of this topic will be the subject of our future study.


\section*{Acknowledgements}
\addcontentsline{toc}{section}{Acknowledgements}

This research was supported by the undergraduate summer research program (REU) of the Physics Department of the Pennsylvania State University.

\section*{Data Availability}

The data underlying this article will be shared on reasonable request to the corresponding author.



\bibliographystyle{mnras}
\bibliography{references} 

\begin{thebibliography}{}
\makeatletter
\relax
\def\mn@urlcharsother{\let\do\@makeother \do\$\do\&\do\#\do\^\do\_\do\%\do\~}
\def\mn@doi{\begingroup\mn@urlcharsother \@ifnextchar [ {\mn@doi@}
  {\mn@doi@[]}}
\def\mn@doi@[#1]#2{\def\@tempa{#1}\ifx\@tempa\@empty \href
  {http://dx.doi.org/#2} {doi:#2}\else \href {http://dx.doi.org/#2} {#1}\fi
  \endgroup}
\def\mn@eprint#1#2{\mn@eprint@#1:#2::\@nil}
\def\mn@eprint@arXiv#1{\href {http://arxiv.org/abs/#1} {{\tt arXiv:#1}}}
\def\mn@eprint@dblp#1{\href {http://dblp.uni-trier.de/rec/bibtex/#1.xml}
  {dblp:#1}}
\def\mn@eprint@#1:#2:#3:#4\@nil{\def\@tempa {#1}\def\@tempb {#2}\def\@tempc
  {#3}\ifx \@tempc \@empty \let \@tempc \@tempb \let \@tempb \@tempa \fi \ifx
  \@tempb \@empty \def\@tempb {arXiv}\fi \@ifundefined
  {mn@eprint@\@tempb}{\@tempb:\@tempc}{\expandafter \expandafter \csname
  mn@eprint@\@tempb\endcsname \expandafter{\@tempc}}}

\bibitem[\protect\citeauthoryear{{Andrews}, {Smith}, {McCully}, {Fox},
  {Valenti}  \& {Howell}}{{Andrews} et~al.}{2017}]{andrews:2017}
{Andrews} J.~E.,  {Smith} N.,  {McCully} C.,  {Fox} O.~D.,  {Valenti} S.,
  {Howell} D.~A.,  2017, \mn@doi [\mnras] {10.1093/mnras/stx1844}, \href
  {https://ui.adsabs.harvard.edu/abs/2017MNRAS.471.4047A} {471, 4047}

\bibitem[\protect\citeauthoryear{{Bersten}, {Benvenuto}  \& {Hamuy}}{{Bersten}
  et~al.}{2011}]{bersten:2011}
{Bersten} M.~C.,  {Benvenuto} O.,   {Hamuy} M.,  2011, \mn@doi [\apj]
  {10.1088/0004-637X/729/1/61}, \href
  {https://ui.adsabs.harvard.edu/abs/2011ApJ...729...61B} {729, 61}

\bibitem[\protect\citeauthoryear{{Bullivant} et~al.,}{{Bullivant}
  et~al.}{2018}]{bullivant:2018}
{Bullivant} C.,  et~al., 2018, \mn@doi [\mnras] {10.1093/mnras/sty045}, \href
  {https://ui.adsabs.harvard.edu/abs/2018MNRAS.476.1497B} {476, 1497}

\bibitem[\protect\citeauthoryear{{Chugai}}{{Chugai}}{1991}]{chugai:1991}
{Chugai} N.~N.,  1991, \mn@doi [\mnras] {10.1093/mnras/250.3.513}, \href
  {https://ui.adsabs.harvard.edu/abs/1991MNRAS.250..513C} {250, 513}

\bibitem[\protect\citeauthoryear{{Chugai}}{{Chugai}}{2001}]{chugai:2001}
{Chugai} N.~N.,  2001, \mn@doi [\mnras] {10.1111/j.1365-2966.2001.04717.x},
  \href {https://ui.adsabs.harvard.edu/abs/2001MNRAS.326.1448C} {326, 1448}

\bibitem[\protect\citeauthoryear{{Chugai} \& {Danziger}}{{Chugai} \&
  {Danziger}}{1994}]{chugai:1994}
{Chugai} N.~N.,  {Danziger} I.~J.,  1994, \mn@doi [\mnras]
  {10.1093/mnras/268.1.173}, \href
  {https://ui.adsabs.harvard.edu/abs/1994MNRAS.268..173C} {268, 173}

\bibitem[\protect\citeauthoryear{{Chugai} et~al.,}{{Chugai}
  et~al.}{2004}]{chugai:2004}
{Chugai} N.~N.,  et~al., 2004, \mn@doi [\mnras]
  {10.1111/j.1365-2966.2004.08011.x}, \href
  {https://ui.adsabs.harvard.edu/abs/2004MNRAS.352.1213C} {352, 1213}

\bibitem[\protect\citeauthoryear{{Das} \& {Ray}}{{Das} \&
  {Ray}}{2017}]{das:2017}
{Das} S.,  {Ray} A.,  2017, \mn@doi [\apj] {10.3847/1538-4357/aa97e1}, \href
  {https://ui.adsabs.harvard.edu/abs/2017ApJ...851..138D} {851, 138}

\bibitem[\protect\citeauthoryear{{Dessart}, {Hillier}, {Audit}, {Livne}  \&
  {Waldman}}{{Dessart} et~al.}{2016}]{dessart:2016}
{Dessart} L.,  {Hillier} D.~J.,  {Audit} E.,  {Livne} E.,   {Waldman} R.,
  2016, \mn@doi [\mnras] {10.1093/mnras/stw336}, \href
  {https://ui.adsabs.harvard.edu/abs/2016MNRAS.458.2094D} {458, 2094}

\bibitem[\protect\citeauthoryear{{Elias-Rosa} et~al.,}{{Elias-Rosa}
  et~al.}{2016}]{elias-rosa:2016}
{Elias-Rosa} N.,  et~al., 2016, \mn@doi [\mnras] {10.1093/mnras/stw2253}, \href
  {https://ui.adsabs.harvard.edu/abs/2016MNRAS.463.3894E} {463, 3894}

\bibitem[\protect\citeauthoryear{{F{\"o}rster} et~al.,}{{F{\"o}rster}
  et~al.}{2018}]{foerster:2018}
{F{\"o}rster} F.,  et~al., 2018, \mn@doi [Nature Astronomy]
  {10.1038/s41550-018-0563-4}, \href
  {https://ui.adsabs.harvard.edu/abs/2018NatAs...2..808F} {2, 808}

\bibitem[\protect\citeauthoryear{{Fox}, {Filippenko}, {Skrutskie}, {Silverman},
  {Ganeshalingam}, {Cenko}  \& {Clubb}}{{Fox} et~al.}{2013}]{fox:2013}
{Fox} O.~D.,  {Filippenko} A.~V.,  {Skrutskie} M.~F.,  {Silverman} J.~M.,
  {Ganeshalingam} M.,  {Cenko} S.~B.,   {Clubb} K.~I.,  2013, \mn@doi [\aj]
  {10.1088/0004-6256/146/1/2}, \href
  {https://ui.adsabs.harvard.edu/abs/2013AJ....146....2F} {146, 2}

\bibitem[\protect\citeauthoryear{{Fraser} et~al.,}{{Fraser}
  et~al.}{2013}]{fraser:2013}
{Fraser} M.,  et~al., 2013, \mn@doi [\apjl] {10.1088/2041-8205/779/1/L8}, \href
  {https://ui.adsabs.harvard.edu/abs/2013ApJ...779L...8F} {779, L8}

\bibitem[\protect\citeauthoryear{{Fuller}}{{Fuller}}{2017}]{Fuller:2017}
{Fuller} J.,  2017, \mn@doi [Monthly Notices of the Royal Astronomical Society]
  {10.1093/mnras/stx1314}, \href
  {https://ui.adsabs.harvard.edu/abs/2017MNRAS.470.1642F} {470, 1642}

\bibitem[\protect\citeauthoryear{{Gal-Yam} et~al.,}{{Gal-Yam}
  et~al.}{2014}]{galyam:2014}
{Gal-Yam} A.,  et~al., 2014, \mn@doi [\nat] {10.1038/nature13304}, \href
  {http://adsabs.harvard.edu/abs/2014Natur.509..471G} {509, 471}

\bibitem[\protect\citeauthoryear{{Grasberg} \& {Nadezhin}}{{Grasberg} \&
  {Nadezhin}}{1986}]{grasberg:1986}
{Grasberg} E.~K.,  {Nadezhin} D.~K.,  1986, Soviet Astronomy Letters, \href
  {https://ui.adsabs.harvard.edu/abs/1986SvAL...12...68G} {12, 68}

\bibitem[\protect\citeauthoryear{{Grasberg} \& {Nadezhin}}{{Grasberg} \&
  {Nadezhin}}{1991}]{grasberg:1991}
{Grasberg} E.~K.,  {Nadezhin} D.~K.,  1991, \sovast, \href
  {https://ui.adsabs.harvard.edu/abs/1991SvA....35...42G} {35, 42}

\bibitem[\protect\citeauthoryear{{Habergham}, {Anderson}, {James}  \&
  {Lyman}}{{Habergham} et~al.}{2014}]{habergham:2014}
{Habergham} S.~M.,  {Anderson} J.~P.,  {James} P.~A.,   {Lyman} J.~D.,  2014,
  \mn@doi [\mnras] {10.1093/mnras/stu684}, \href
  {https://ui.adsabs.harvard.edu/abs/2014MNRAS.441.2230H} {441, 2230}

\bibitem[\protect\citeauthoryear{{Hosseinzadeh} et~al.,}{{Hosseinzadeh}
  et~al.}{2018}]{hosseinzadeh:2018}
{Hosseinzadeh} G.,  et~al., 2018, \mn@doi [\apj] {10.3847/1538-4357/aac5f6},
  \href {https://ui.adsabs.harvard.edu/abs/2018ApJ...861...63H} {861, 63}

\bibitem[\protect\citeauthoryear{{Johnson}, {Kochanek}  \& {Adams}}{{Johnson}
  et~al.}{2018}]{johnson:2017}
{Johnson} S.~A.,  {Kochanek} C.~S.,   {Adams} S.~M.,  2018, \mn@doi [\mnras]
  {10.1093/mnras/sty1966}, \href
  {http://adsabs.harvard.edu/abs/2018MNRAS.480.1696J} {480, 1696}

\bibitem[\protect\citeauthoryear{{Kasen} \& {Woosley}}{{Kasen} \&
  {Woosley}}{2009}]{kasen:2009}
{Kasen} D.,  {Woosley} S.~E.,  2009, \mn@doi [\apj]
  {10.1088/0004-637X/703/2/2205}, \href
  {https://ui.adsabs.harvard.edu/abs/2009ApJ...703.2205K} {703, 2205}

\bibitem[\protect\citeauthoryear{{Khazov} et~al.,}{{Khazov}
  et~al.}{2016}]{khazov:2016}
{Khazov} D.,  et~al., 2016, \mn@doi [\apj] {10.3847/0004-637X/818/1/3}, \href
  {http://adsabs.harvard.edu/abs/2016ApJ...818....3K} {818, 3}

\bibitem[\protect\citeauthoryear{{Kiewe} et~al.,}{{Kiewe}
  et~al.}{2012}]{kiewe:2012}
{Kiewe} M.,  et~al., 2012, \mn@doi [\apj] {10.1088/0004-637X/744/1/10}, \href
  {https://ui.adsabs.harvard.edu/abs/2012ApJ...744...10K} {744, 10}

\bibitem[\protect\citeauthoryear{{Kochanek} et~al.,}{{Kochanek}
  et~al.}{2017}]{kochanek:2017}
{Kochanek} C.~S.,  et~al., 2017, \mn@doi [\mnras] {10.1093/mnras/stx291}, \href
  {http://adsabs.harvard.edu/abs/2017MNRAS.467.3347K} {467, 3347}

\bibitem[\protect\citeauthoryear{{Kurf{\"u}rst}, {Pejcha}  \&
  {Krti{\v{c}}ka}}{{Kurf{\"u}rst} et~al.}{2020}]{kurfurst:2020}
{Kurf{\"u}rst} P.,  {Pejcha} O.,   {Krti{\v{c}}ka} J.,  2020, \mn@doi [\aap]
  {10.1051/0004-6361/202039073}, \href
  {https://ui.adsabs.harvard.edu/abs/2020A&A...642A.214K} {642, A214}

\bibitem[\protect\citeauthoryear{{Leung} \& {Fuller}}{{Leung} \&
  {Fuller}}{2020}]{leung:2020}
{Leung} S.-C.,  {Fuller} J.,  2020, \mn@doi [The Astrophysical Journal]
  {10.3847/1538-4357/abac5d}, \href
  {https://ui.adsabs.harvard.edu/abs/2020ApJ...900...99L} {900, 99}

\bibitem[\protect\citeauthoryear{{Lin} et~al.,}{{Lin} et~al.}{2021}]{lin:2021}
{Lin} H.,  et~al., 2021, \mn@doi [\mnras] {10.1093/mnras/stab1550}, \href
  {https://ui.adsabs.harvard.edu/abs/2021MNRAS.505.4890L} {505, 4890}

\bibitem[\protect\citeauthoryear{{Litvinova} \& {Nadezhin}}{{Litvinova} \&
  {Nadezhin}}{1983}]{litvinova:1983}
{Litvinova} I.~I.,  {Nadezhin} D.~K.,  1983, \mn@doi [\apss]
  {10.1007/BF01008387}, \href
  {https://ui.adsabs.harvard.edu/abs/1983Ap&SS..89...89L} {89, 89}

\bibitem[\protect\citeauthoryear{{Margutti} et~al.,}{{Margutti}
  et~al.}{2014}]{margutti:2014}
{Margutti} R.,  et~al., 2014, \mn@doi [\apj] {10.1088/0004-637X/780/1/21},
  \href {https://ui.adsabs.harvard.edu/abs/2014ApJ...780...21M} {780, 21}

\bibitem[\protect\citeauthoryear{{Mauerhan} et~al.,}{{Mauerhan}
  et~al.}{2013a}]{mauerhan:2013}
{Mauerhan} J.~C.,  et~al., 2013a, \mn@doi [\mnras] {10.1093/mnras/stt009},
  \href {https://ui.adsabs.harvard.edu/abs/2013MNRAS.430.1801M} {430, 1801}

\bibitem[\protect\citeauthoryear{{Mauerhan} et~al.,}{{Mauerhan}
  et~al.}{2013b}]{mauerhan:2013a}
{Mauerhan} J.~C.,  et~al., 2013b, \mn@doi [\mnras] {10.1093/mnras/stt360},
  \href {https://ui.adsabs.harvard.edu/abs/2013MNRAS.431.2599M} {431, 2599}

\bibitem[\protect\citeauthoryear{{Moriya}}{{Moriya}}{2014}]{moriya:2014}
{Moriya} T.~J.,  2014, \mn@doi [\aap] {10.1051/0004-6361/201322992}, \href
  {https://ui.adsabs.harvard.edu/abs/2014A&A...564A..83M} {564, A83}

\bibitem[\protect\citeauthoryear{{Moriya}, {Maeda}, {Taddia}, {Sollerman},
  {Blinnikov}  \& {Sorokina}}{{Moriya} et~al.}{2014a}]{moriya:2014a}
{Moriya} T.~J.,  {Maeda} K.,  {Taddia} F.,  {Sollerman} J.,  {Blinnikov} S.~I.,
    {Sorokina} E.~I.,  2014a, \mn@doi [\mnras] {10.1093/mnras/stu163}, \href
  {https://ui.adsabs.harvard.edu/abs/2014MNRAS.439.2917M} {439, 2917}

\bibitem[\protect\citeauthoryear{{Moriya}, {Tominaga}, {Langer}, {Nomoto},
  {Blinnikov}  \& {Sorokina}}{{Moriya} et~al.}{2014b}]{moriya:2014b}
{Moriya} T.~J.,  {Tominaga} N.,  {Langer} N.,  {Nomoto} K.,  {Blinnikov} S.~I.,
    {Sorokina} E.~I.,  2014b, \mn@doi [\aap] {10.1051/0004-6361/201424264},
  \href {https://ui.adsabs.harvard.edu/abs/2014A&A...569A..57M} {569, A57}

\bibitem[\protect\citeauthoryear{{Moriya}, {F{\"o}rster}, {Yoon},
  {Gr{\"a}fener}  \& {Blinnikov}}{{Moriya} et~al.}{2018}]{moriya:2018}
{Moriya} T.~J.,  {F{\"o}rster} F.,  {Yoon} S.-C.,  {Gr{\"a}fener} G.,
  {Blinnikov} S.~I.,  2018, \mn@doi [\mnras] {10.1093/mnras/sty475}, \href
  {http://adsabs.harvard.edu/abs/2018MNRAS.tmp..463M} {}

\bibitem[\protect\citeauthoryear{{Morozova}, {Piro}, {Renzo}, {Ott}, {Clausen},
  {Couch}, {Ellis}  \& {Roberts}}{{Morozova} et~al.}{2015}]{Morozova:2015}
{Morozova} V.,  {Piro} A.~L.,  {Renzo} M.,  {Ott} C.~D.,  {Clausen} D.,
  {Couch} S.~M.,  {Ellis} J.,   {Roberts} L.~F.,  2015, \mn@doi [\apj]
  {10.1088/0004-637X/814/1/63}, \href
  {https://ui.adsabs.harvard.edu/abs/2015ApJ...814...63M} {814, 63}

\bibitem[\protect\citeauthoryear{{Morozova}, {Piro}  \& {Valenti}}{{Morozova}
  et~al.}{2017}]{morozova:2017}
{Morozova} V.,  {Piro} A.~L.,   {Valenti} S.,  2017, \mn@doi [\apj]
  {10.3847/1538-4357/aa6251}, \href
  {http://adsabs.harvard.edu/abs/2017ApJ...838...28M} {838, 28}

\bibitem[\protect\citeauthoryear{{Morozova}, {Piro}  \& {Valenti}}{{Morozova}
  et~al.}{2018}]{morozova:2017a}
{Morozova} V.,  {Piro} A.~L.,   {Valenti} S.,  2018, \mn@doi [\apj]
  {10.3847/1538-4357/aab9a6}, \href
  {https://ui.adsabs.harvard.edu/abs/2018ApJ...858...15M} {858, 15}

\bibitem[\protect\citeauthoryear{{Morozova}, {Piro}, {Fuller}  \& {Van
  Dyk}}{{Morozova} et~al.}{2020}]{morozova:2020}
{Morozova} V.,  {Piro} A.~L.,  {Fuller} J.,   {Van Dyk} S.~D.,  2020, \mn@doi
  [The Astrophysical Journal Letters] {10.3847/2041-8213/ab77c8}, \href
  {https://ui.adsabs.harvard.edu/abs/2020ApJ...891L..32M} {891, L32}

\bibitem[\protect\citeauthoryear{{Nagy} \& {Vink{\'o}}}{{Nagy} \&
  {Vink{\'o}}}{2016}]{nagy:2016}
{Nagy} A.~P.,  {Vink{\'o}} J.,  2016, \mn@doi [\aap]
  {10.1051/0004-6361/201527931}, \href
  {http://adsabs.harvard.edu/abs/2016A%26A...589A..53N} {589, A53}

\bibitem[\protect\citeauthoryear{{Nakaoka} et~al.,}{{Nakaoka}
  et~al.}{2018}]{nakaoka:2018}
{Nakaoka} T.,  et~al., 2018, \mn@doi [\apj] {10.3847/1538-4357/aabee7}, \href
  {http://adsabs.harvard.edu/abs/2018ApJ...859...78N} {859, 78}

\bibitem[\protect\citeauthoryear{{Nyholm} et~al.,}{{Nyholm}
  et~al.}{2017}]{nyholm:2017}
{Nyholm} A.,  et~al., 2017, \mn@doi [\aap] {10.1051/0004-6361/201629906}, \href
  {https://ui.adsabs.harvard.edu/abs/2017A&A...605A...6N} {605, A6}

\bibitem[\protect\citeauthoryear{{Nyholm} et~al.,}{{Nyholm}
  et~al.}{2020}]{nyholm:2020}
{Nyholm} A.,  et~al., 2020, \mn@doi [\aap] {10.1051/0004-6361/201936097}, \href
  {https://ui.adsabs.harvard.edu/abs/2020A&A...637A..73N} {637, A73}

\bibitem[\protect\citeauthoryear{{O'Neill} et~al.,}{{O'Neill}
  et~al.}{2019}]{oneill:2018}
{O'Neill} D.,  et~al., 2019, \mn@doi [\aap] {10.1051/0004-6361/201834566},
  \href {https://ui.adsabs.harvard.edu/abs/2019A&A...622L...1O} {622, L1}

\bibitem[\protect\citeauthoryear{{Ofek} et~al.,}{{Ofek}
  et~al.}{2013}]{ofek:2013}
{Ofek} E.~O.,  et~al., 2013, \mn@doi [\nat] {10.1038/nature11877}, \href
  {https://ui.adsabs.harvard.edu/abs/2013Natur.494...65O} {494, 65}

\bibitem[\protect\citeauthoryear{{Ofek} et~al.,}{{Ofek}
  et~al.}{2014}]{ofek:2014}
{Ofek} E.~O.,  et~al., 2014, \mn@doi [\apj] {10.1088/0004-637X/789/2/104},
  \href {https://ui.adsabs.harvard.edu/abs/2014ApJ...789..104O} {789, 104}

\bibitem[\protect\citeauthoryear{{Ofek} et~al.,}{{Ofek}
  et~al.}{2016}]{ofek:2016}
{Ofek} E.~O.,  et~al., 2016, \mn@doi [\apj] {10.3847/0004-637X/824/1/6}, \href
  {https://ui.adsabs.harvard.edu/abs/2016ApJ...824....6O} {824, 6}

\bibitem[\protect\citeauthoryear{{Pastorello} et~al.,}{{Pastorello}
  et~al.}{2018}]{pastorello:2018}
{Pastorello} A.,  et~al., 2018, \mn@doi [\mnras] {10.1093/mnras/stx2668}, \href
  {https://ui.adsabs.harvard.edu/abs/2018MNRAS.474..197P} {474, 197}

\bibitem[\protect\citeauthoryear{{Paxton} et~al.,}{{Paxton}
  et~al.}{2018}]{paxton:2018}
{Paxton} B.,  et~al., 2018, \mn@doi [\apjs] {10.3847/1538-4365/aaa5a8}, \href
  {http://adsabs.harvard.edu/abs/2018ApJS..234...34P} {234, 34}

\bibitem[\protect\citeauthoryear{{Prieto} et~al.,}{{Prieto}
  et~al.}{2008}]{prieto:2008}
{Prieto} J.~L.,  et~al., 2008, \mn@doi [\apjl] {10.1086/589922}, \href
  {https://ui.adsabs.harvard.edu/abs/2008ApJ...681L...9P} {681, L9}

\bibitem[\protect\citeauthoryear{{Pumo}, {Zampieri}, {Spiro}, {Pastorello},
  {Benetti}, {Cappellaro}, {Manic{\`o}}  \& {Turatto}}{{Pumo}
  et~al.}{2017}]{pumo:2017}
{Pumo} M.~L.,  {Zampieri} L.,  {Spiro} S.,  {Pastorello} A.,  {Benetti} S.,
  {Cappellaro} E.,  {Manic{\`o}} G.,   {Turatto} M.,  2017, \mn@doi [\mnras]
  {10.1093/mnras/stw2625}, \href
  {http://adsabs.harvard.edu/abs/2017MNRAS.464.3013P} {464, 3013}

\bibitem[\protect\citeauthoryear{{Quataert} \& {Shiode}}{{Quataert} \&
  {Shiode}}{2012}]{quataert:2012}
{Quataert} E.,  {Shiode} J.,  2012, \mn@doi [\mnras]
  {10.1111/j.1745-3933.2012.01264.x}, \href
  {https://ui.adsabs.harvard.edu/abs/2012MNRAS.423L..92Q} {423, L92}

\bibitem[\protect\citeauthoryear{{Reguitti} et~al.,}{{Reguitti}
  et~al.}{2019}]{reguitti:2019}
{Reguitti} A.,  et~al., 2019, \mn@doi [\mnras] {10.1093/mnras/sty2870}, \href
  {https://ui.adsabs.harvard.edu/abs/2019MNRAS.482.2750R} {482, 2750}

\bibitem[\protect\citeauthoryear{{Schlegel}}{{Schlegel}}{1990}]{schlegel:1990}
{Schlegel} E.~M.,  1990, \mnras, \href
  {https://ui.adsabs.harvard.edu/abs/1990MNRAS.244..269S} {244, 269}

\bibitem[\protect\citeauthoryear{{Shiode} \& {Quataert}}{{Shiode} \&
  {Quataert}}{2014}]{shiode:2014}
{Shiode} J.~H.,  {Quataert} E.,  2014, \mn@doi [\apj]
  {10.1088/0004-637X/780/1/96}, \href
  {https://ui.adsabs.harvard.edu/abs/2014ApJ...780...96S} {780, 96}

\bibitem[\protect\citeauthoryear{{Smith}}{{Smith}}{2017}]{smith:2017}
{Smith} N.,  2017, {Interacting Supernovae: Types IIn and Ibn}.
p.~403, \mn@doi{10.1007/978-3-319-21846-5\_38}

\bibitem[\protect\citeauthoryear{{Smith} et~al.,}{{Smith}
  et~al.}{2015}]{smith:2015}
{Smith} N.,  et~al., 2015, \mn@doi [\mnras] {10.1093/mnras/stv354}, \href
  {http://adsabs.harvard.edu/abs/2015MNRAS.449.1876S} {449, 1876}

\bibitem[\protect\citeauthoryear{{Strotjohann} et~al.,}{{Strotjohann}
  et~al.}{2021}]{strotjohann:2021}
{Strotjohann} N.~L.,  et~al., 2021, \mn@doi [\apj] {10.3847/1538-4357/abd032},
  \href {https://ui.adsabs.harvard.edu/abs/2021ApJ...907...99S} {907, 99}

\bibitem[\protect\citeauthoryear{{Sukhbold}, {Ertl}, {Woosley}, {Brown}  \&
  {Janka}}{{Sukhbold} et~al.}{2016}]{sukhbold:2016}
{Sukhbold} T.,  {Ertl} T.,  {Woosley} S.~E.,  {Brown} J.~M.,   {Janka} H.~T.,
  2016, \mn@doi [\apj] {10.3847/0004-637X/821/1/38}, \href
  {https://ui.adsabs.harvard.edu/abs/2016ApJ...821...38S} {821, 38}

\bibitem[\protect\citeauthoryear{{Szczygie{\l}}, {Kochanek}  \&
  {Dai}}{{Szczygie{\l}} et~al.}{2012}]{szczygiel:2012}
{Szczygie{\l}} D.~M.,  {Kochanek} C.~S.,   {Dai} X.,  2012, \mn@doi [\apj]
  {10.1088/0004-637X/760/1/20}, \href
  {https://ui.adsabs.harvard.edu/abs/2012ApJ...760...20S} {760, 20}

\bibitem[\protect\citeauthoryear{{Taddia} et~al.,}{{Taddia}
  et~al.}{2013}]{taddia:2013}
{Taddia} F.,  et~al., 2013, \mn@doi [\aap] {10.1051/0004-6361/201321180}, \href
  {https://ui.adsabs.harvard.edu/abs/2013A&A...555A..10T} {555, A10}

\bibitem[\protect\citeauthoryear{{Takei}, {Tsuna}, {Kuriyama}, {Ko}  \&
  {Shigeyama}}{{Takei} et~al.}{2021}]{takei:2021}
{Takei} Y.,  {Tsuna} D.,  {Kuriyama} N.,  {Ko} T.,   {Shigeyama} T.,  2021,
  arXiv e-prints, \href {https://ui.adsabs.harvard.edu/abs/2021arXiv210905871T}
  {p. arXiv:2109.05871}

\bibitem[\protect\citeauthoryear{{Tartaglia} et~al.,}{{Tartaglia}
  et~al.}{2016}]{tartaglia:2016}
{Tartaglia} L.,  et~al., 2016, \mn@doi [\mnras] {10.1093/mnras/stw675}, \href
  {https://ui.adsabs.harvard.edu/abs/2016MNRAS.459.1039T} {459, 1039}

\bibitem[\protect\citeauthoryear{{Th{\"o}ne} et~al.,}{{Th{\"o}ne}
  et~al.}{2017}]{thoene:2017}
{Th{\"o}ne} C.~C.,  et~al., 2017, \mn@doi [\aap] {10.1051/0004-6361/201629968},
  \href {https://ui.adsabs.harvard.edu/abs/2017A&A...599A.129T} {599, A129}

\bibitem[\protect\citeauthoryear{{Tinyanont} et~al.,}{{Tinyanont}
  et~al.}{2021}]{tinyanont:2021}
{Tinyanont} S.,  et~al., 2021, \mn@doi [\mnras] {10.1093/mnras/stab2887}, \href
  {https://ui.adsabs.harvard.edu/abs/2021MNRAS.tmp.2821T} {}

\bibitem[\protect\citeauthoryear{{Valenti} et~al.,}{{Valenti}
  et~al.}{2016}]{valenti:2016}
{Valenti} S.,  et~al., 2016, \mn@doi [\mnras] {10.1093/mnras/stw870}, \href
  {https://ui.adsabs.harvard.edu/abs/2016MNRAS.459.3939V} {459, 3939}

\bibitem[\protect\citeauthoryear{{Wu} \& {Fuller}}{{Wu} \&
  {Fuller}}{2021}]{wu:2021}
{Wu} S.,  {Fuller} J.,  2021, \mn@doi [The Astrophysical Journal]
  {10.3847/1538-4357/abc87c}, \href
  {https://ui.adsabs.harvard.edu/abs/2021ApJ...906....3W} {906, 3}

\bibitem[\protect\citeauthoryear{{Yaron} et~al.,}{{Yaron}
  et~al.}{2017}]{yaron:2017}
{Yaron} O.,  et~al., 2017, \mn@doi [Nature Physics] {10.1038/nphys4025}, \href
  {http://adsabs.harvard.edu/abs/2017NatPh..13..510Y} {13, 510}

\bibitem[\protect\citeauthoryear{{de la Rosa}, {Roming}, {Pritchard}  \&
  {Fryer}}{{de la Rosa} et~al.}{2016}]{delarosa:2016}
{de la Rosa} J.,  {Roming} P.,  {Pritchard} T.,   {Fryer} C.,  2016, \mn@doi
  [\apj] {10.3847/0004-637X/820/1/74}, \href
  {https://ui.adsabs.harvard.edu/abs/2016ApJ...820...74D} {820, 74}

\makeatother
\end{thebibliography}





\bsp	
\label{lastpage}
\end{document}